\documentclass[manuscript,screen,nonacm]{acmart}
\AtBeginDocument{%
  }

\setcopyright{none}    

\makeatletter
\renewcommand\footnotetextcopyrightpermission[1]{} % kills "Manuscript submitted to ACM"
\makeatother

\authorsaddresses{}

\usepackage{algorithm}
\usepackage{algpseudocode}
\usepackage{graphicx}
\usepackage{subcaption}

\begin{document}

\title{Multi-Agent Code-Orchestrated Generation for Reliable Infrastructure-as-Code}


\author{Rana Nameer Hussain Khan}
\email{rnameerkhan@vt.edu}
\affiliation{%
  \institution{Virginia Tech}
  \city{Blacksburg}
  \state{VA}
  \country{USA}
}

\author{Dawood Wasif}
\email{dawoodwasif@vt.edu}
\affiliation{%
  \institution{Virginia Tech}
  \city{Blacksburg}
  \state{VA}
  \country{USA}
}

\author{Jin-Hee Cho}
\email{jicho@vt.edu}
\affiliation{%
  \institution{Virginia Tech}
  \city{Blacksburg}
  \state{VA}
  \country{USA}
}

\author{Ali Butt}
\email{butta@vt.edu}
\affiliation{%
  \institution{Virginia Tech}
  \city{Blacksburg}
  \state{VA}
  \country{USA}
}

%%
%% By default, the full list of authors will be used on the page
%% headers. Often, this list is too long and will overlap
\renewcommand{\shortauthors}{Khan et al.}

\begin{abstract}
  The increasing complexity of cloud-native infrastructure has made Infrastructure-as-Code (IaC) essential for reproducible and scalable deployments. While large language models (LLMs) have shown promise in generating IaC snippets from natural language prompts, their monolithic, single-pass generation approach often results in syntactic errors, policy violations, and unscalable designs. In this paper, we propose MACOG (Multi-Agent Code-Orchestrated Generation), a novel multi-agent LLM-based architecture for IaC generation that decomposes the task into modular subtasks handled by specialized agents: Architect, Provider Harmonizer, Engineer, Reviewer, Security Prover, Cost and Capacity Planner, DevOps, and Memory Curator. The agents interact via a shared-blackboard, finite-state orchestrator layer, and collectively produce Terraform configurations that are not only syntactically valid but also policy-compliant and semantically coherent. To ensure infrastructure correctness and governance, we incorporate Terraform Plan for execution validation and Open Policy Agent (OPA) for customizable policy enforcement. We evaluate MACOG using the IaC-Eval benchmark, where MACOG is the top enhancement across models, e.g., GPT-5 improves from 54.90 (RAG) to 74.02 and Gemini-2.5 Pro from 43.56 to 60.13, with concurrent gains on BLEU, CodeBERTScore, and an LLM-judge metric. Ablations show constrained decoding and deploy feedback are critical: removing them drops IaC-Eval to 64.89 and 56.93, respectively.
\end{abstract}

% --- CCS Concepts (concise; no XML during review) ---

% --- Keywords (5–7 concise phrases) ---



%%
%% This command processes the author and affiliation, and title
\maketitle

% Abstract -> at end
% Introduction (Rana)
% Related Works (Rana)
% Methodology (Dawood)
% Experimental Setup
% Experimental Results
% Discussion and Limitations
% Conclusions

\section{Introduction}

Modern cloud platforms expose a rich and evolving surface of services, configuration knobs, and compliance regimes, and Infrastructure-as-Code (IaC) \cite{rahman2019systematic} has become the common medium that teams use to tame this complexity. Terraform, Pulumi, and CloudFormation encode desired state as declarative programs that must be valid with respect to provider schemas, consistent across interdependent resources, and faithful to organizational rules on security, cost, and data residency. Large language models (LLMs) \cite{chang2024survey} promise to shorten the distance from a plain-English specification to a working IaC program, but the path from intent to a deployable, auditable artifact is fraught with domain-specific traps: strict schemas with versioned fields, cross-resource references with subtle naming constraints, non-obvious dependency orderings, and environment-specific quirks that only surface at \texttt{plan/apply} time. Beyond mere syntax, production-grade configurations must satisfy non-functional expectations such as least-privilege access, encryption at rest and in transit, region pinning for residency, redundancy for availability targets, and budget ceilings. The result is a synthesis task that is not just code generation, but constrained program construction under multiple interacting validators that each speak a different dialect of evidence, from static type checks to policy proofs and runtime logs

The operational setting amplifies these difficulties. Provider schemas change, default values shift, and resource classes deprecate or migrate across regions, creating a moving target for any static set of examples. Teams often assemble configurations by composing modules that were written months apart under different assumptions, causing subtle mismatches in variable interfaces and output names that defeat trivial pattern matching. Additionally, IaC is inherently graph-shaped: a VPC frames subnets, gateways, and route tables; security groups define ingress and egress edges that must line up with compute nodes and managed databases; identity and access policies must name concrete ARNs that exist only after other resources are planned. This graph structure is not an incidental detail but the main act, and language models that treat code as flat text frequently stumble on global constraints such as acyclicity, topological ordering, and cross-file coherency. When models emit a near-correct configuration, the last mile still tends to break on details such as \texttt{subnet} versus \texttt{subnet\_id}, reference scoping across modules, or region-specific capabilities. Finally, non-functional constraints do not present a single, unified interface: cost is numeric and region-aware, security policies are logical formulas evaluated by engines like OPA or scanners such as Checkov/Regula, and deployability depends on real toolchains executing in realistic sandboxes. Any practical assistant must place these validators at the center of the workflow rather than treat them as afterthoughts

Recent efforts have aimed at closing this gap with prompting and retrieval strategies. Few-shot prompting improves local idioms and reduces obvious syntax errors, but scales poorly when intents span multiple providers, when the plan includes three or more interlocked modules, or when token budgets force elision of the very context that would disambiguate a reference. Retrieval-augmented approaches can surface nearby examples, yet raw snippets are brittle: a single field that changed between provider versions derails an otherwise promising candidate, and unaudited retrieval cannot guarantee that a borrowed pattern observes the organization’s policies. Single-agent, multi-turn scaffolds with tool calls can iterate on errors, though they tend to oscillate in long contexts, overwrite working parts during late repairs, and struggle to keep a coherent, typed view of the infrastructure graph over many steps. More structured variants propose plan-then-code or JSON-first pipelines, but without a typed intermediate representation and grammar-aware decoding, the realization step reintroduces inconsistencies, and without tight feedback from external validators, the loop lacks the counterexamples needed for surgical repairs. Crucially, many systems rely on fine-tuning to imprint domain behavior; that path is costly to maintain across providers and versions and often under-delivers in the face of real \texttt{plan/apply} idiosyncrasies

This paper takes a different route by organizing the task around the validators and the graph structure of the infrastructure. We present a method that keeps large language models in an instruction-following, zero fine-tune mode and surrounds them with strict structure, deterministic compilation, grammar- and schema-constrained decoding, and an error-driven repair loop. A typed Infrastructure Intermediate Representation (I-IR) serves as the shared language between agents and tools, making resources, edges, regions, and effects explicit and checkable. The workflow compiles I-IR into Terraform through resource skeletons and a constrained decoder that cannot stray outside the HCL grammar \cite{riti2021beginning} or provider field sets, enforces a round-trip check back into I-IR to preserve intent, and then subjects the candidate to a battery of validators: \texttt{terraform validate} for schema and references, OPA/Rego and complementary scanners for policy, deterministic price books for budget, and realistic sandboxes for deployment. Counterexamples from any stage are mapped to minimal plan-level or code-level edits, reducing failures without thrashing working parts. A small set of role-specialized agents coordinate over a versioned blackboard and reuse verified \emph{motifs}—typed, provider-versioned plan fragments—in place of raw code snippets. We introduce this approach, \emph{Multi-Agent Code-Orchestrated Generation} (MACOG), at the end of the pipeline for a single, clear purpose: to transform natural-language intents into deployable, compliant, and cost-aware Terraform programs, with an evidence bundle that a third party can verify offline before execution.

\section{Background}

\subsection{DevOps Automation and Evolution}
DevOps is an umbrella term that defines the shift towards high automation and tight integration of software design and infrastructure. Since its inception, DevOps has driven significant changes in the IT world, notably the adoption of practices such as continuous integration and delivery, as well as Infrastructure as Code. Over the years, the degree of automation in these processes has steadily increased. Eventually, a \textbf{Everything as Code} approach became increasingly adopted in the software world. Not only infrastructure, but entire build pipelines, configuration files, and even monitoring checks are being built using code \cite{vakhula2025research}. 

A key enabler in the evolution of DevOps was the advent of containerization. That was mainly spearheaded by the development of Docker \cite{merkel2014docker} and Kubernetes. The shift led to an improvement in accuracy across production environments, but highlighted the need for governance. As deployments scaled up exponentially to hundreds of microservices, manually enforcing compliance and best practices became untenable. To address this, the concept of Policy-as-Code \cite{chittala2024securing} was developed. Policies are expressed in code and are automatically checked at runtime or in CI pipelines. For example, a policy might declare that no AWS S3 bucket should be publicly readable; using PaC frameworks (e.g., Open Policy Agent or HashiCorp Sentinel), such rules are evaluated on IaC changes and block non-compliant infrastructure changes. PaC has become a vital part of DevSecOps, allowing “shift-left” enforcement of security and compliance early in the pipeline.

However, automating the generation of correct infrastructure code is a non-trivial next step. Crafting Terraform or CloudFormation scripts still requires considerable expertise in cloud services and syntax. Unlike application code – for which testing and specification techniques are more mature – infrastructure code deals with real-world configurations that must align with implicit operational requirements (scalability, security, cost-effectiveness). The consequence is a high barrier to entry and a propensity for errors or suboptimal setups in IaC. These pain points are driving interest in intelligent automation: using AI to assist or even fully automate IaC script creation and validation. The evolution of DevOps thus naturally leads to the question: can we apply \textit{learning-based automation} (specifically, Large Language Models) to generate reliable infrastructure code, thereby reducing human toil and error? The background above illustrates both the opportunity (the rich automation and validation ecosystem to build upon) and the need (the difficulty and importance of getting IaC right).

\subsection{Large Language Models for Code Generation in Software Engineering}
The past few years have seen an exponential growth in Large Language Models and their use cases, particularly for code generation \cite{dong2025survey}. Unlike earlier program synthesis approaches that required formal specifications or symbolic reasoning, modern LLMs learn to generate code by statistically modeling massive code corpora. OpenAI’s GPT family and related transformer-based models (e.g., CodeBERT, CodeT5, etc.) have demonstrated the ability to produce syntactically correct and often functional code in a variety of languages. DeepMind’s AlphaCode pushed the frontier further by generating thousands of candidate programs for competitive programming problems and selecting correct ones via test execution, achieving a top 54.3\% ranking on Codeforces challenges \cite{li2022competition}. This shows that when given the right data, LLMs can produce high-quality code capable of solving advanced and complex problems. Crucially in the DeepMind paper \cite{li2022competition}, they also highlighted the importance of coupling generation with validation: AlphaCode’s success hinged on running generated programs against tests and filtering out failures. 

In the domain of software configuration and DevOps, early anecdotal evidence showed models like GPT-4 can produce YAML or Terraform snippets from plain English descriptions; however, achieving correct and safe infrastructure configurations via one-shot generation is challenging. Recent research confirms this gap: Kon et al. (2024) \cite{kon2024iac} introduced IaC-Eval, a benchmark of 458 cloud infrastructure specification tasks, and found that even state-of-the-art LLMs (GPT-4) solved only ~19\% of tasks correctly on the first try. The errors often stem from the model’s hallucinations or lack of contextual understanding of cloud services – for example, missing required resource properties, using incorrect identifiers, or violating cloud-specific constraints (e.g., naming conventions, region restrictions). These results highlight that vanilla LLM generation is not yet reliable for IaC. To bridge this reliability gap, researchers are turning to multi-step, feedback-driven approaches. 

A promising direction is the use of multi-agent or iterative LLM pipelines for code generation. Instead of a single-pass answer, the process is structured into roles or stages that mimic a software team’s workflow: requirements analysis, coding, and testing. Dong et al. (2024) \cite{dong2024self} present a self-collaboration framework where a single ChatGPT instance iteratively plays different roles (e.g., user, coder, tester) in sequence, incorporating feedback at each stage to refine the output. Similarly, Qian et al. (2023) \cite{qian2023chatdev} introduce ChatDev, in which multiple specialized LLM-based agents (e.g., an “Architect”, “Developer”, and “Tester” agent) communicate with each other in natural language to build and verify a program progressively. ChatDev’s agents exchange design ideas, code patches, and test results in a chat chain, successfully developing non-trivial software with minimal human input. Notably, this approach reduced coding errors by having the Tester agent catch failures and prompt the Developer agent to fix them, thereby addressing the hallucination and oversight problems common in single-step generation. These multi-agent systems demonstrate that incorporating an automated feedback loop – especially testing and error correction – can substantially improve the correctness and completeness of generated code. In essence, the LLMs are used not just as code generators, but also as critics and debuggers for each other’s outputs, guided by a predefined collaboration protocol or “playbook” of roles (often inspired by the software development life cycle).

Our work builds on a solid foundation of software engineering research and cutting-edge AI techniques: from the lessons of IaC quality and testing research, we inherit the necessity of policy compliance and verification; from the state-of-the-art in LLM-based code generation, we adopt multi-agent collaboration to improve reliability.

\section{Related Work}
\subsection{LLMs for Infrastructure-as-Code and Configuration Synthesis}
Work on LLMs for configuration domains (Terraform\cite{shirinkin2017getting}, CloudFormation \cite{tovmasyan2020mastering}, Ansible \cite{mcallister2017implementing}) typically frames the task as mapping a natural-language intent into a deployable artifact while satisfying strict schemas and cross-resource references. Empirical studies show one-shot prompting is brittle in IaC: even strong models underperform without tool feedback, often omitting required fields, misusing identifiers, or violating provider constraints \cite{kon2024iac}. Parallel streams in software configuration quality document configuration ``smells'' and anti-patterns that degrade maintainability and security, reinforcing that surface-level correctness is insufficient \cite{sharma2016does,schwarz2018code}. Security-focused analyses catalog recurring IaC smells (``Seven Sins'') such as hard-coded secrets and overly permissive policies, underscoring the need for automated, policy-aware checks \cite{opara2025chaos}. Recent measurements of Terraform security practices across open-source projects further highlight gaps in adoption and enforcement \cite{verdet2023exploring}. Together, these findings motivate IaC methods that combine generation with schema awareness, policy compliance, and runtime validation rather than relying solely on retrieval or few-shot exemplars

\subsection{Multi-Agent and Tool-Augmented Code Generation}
A complementary thread organizes code generation as a collaboration among specialized agents (planner, developer, tester) and external tools. Multi-agent systems such as \emph{ChatDev} demonstrate that role specialization, shared memory, and iterative critique can reduce hallucinations and improve functional correctness on end-to-end software tasks \cite{qian2023chatdev}. \emph{Self-collaboration} shows similar gains by letting one model play multiple roles across plan–code–test cycles \cite{dong2024self}. Beyond general coding, multi-agent frameworks targeted at software evolution (\emph{MAGIS}) coordinate planning and QA to resolve GitHub issues more reliably than single-agent prompting \cite{tao2024magis}. In program repair, agentic designs (\emph{RepairAgent}) couple LLMs with a finite-state tool controller to gather diagnostics, apply patches, and validate fixes autonomously \cite{bouzenia2024repairagent}. These systems share a pattern: structured division of labor plus tool-grounding yields more robust iterations than free-form chat. For IaC, this suggests orchestrations that pass typed artifacts through validators (schema, policy, runtime), enabling precise, minimal edits while preserving previously correct structure

\subsection{Constrained Decoding and Validator-Guided Repair}
Constrained decoding narrows the output space to grammar- or schema-admissible tokens, markedly reducing syntactic invalidity in structured code generation. \emph{PICARD} enforces incremental parsing constraints during decoding to keep outputs valid for formal languages like SQL \cite{scholak2021picard}. More recent methods (\emph{SynCode}) precompute DFA-based masks for CFGs to ensure syntactic validity efficiently across languages \cite{ugare2024syncode}, while \emph{Grammar-Aligned Decoding} (ASAp) addresses distributional bias introduced by hard constraints, aligning sampling with the LLM’s conditional distribution under a grammar \cite{park2024grammar}. On the repair side, classic \emph{counterexample-guided} loops (CEGIS) alternate synthesis with verification, using failed obligations to steer minimal edits \cite{solar2009sketching}. Contemporary LLM repair work formalizes refinement as an exploration–exploitation problem over failing tests and partial successes \cite{tang2024code}. For IaC, combining grammar/schema-constrained realization with validator-guided repair (static checks, policy engines, sandboxed plan/apply) operationalizes these principles: keep generation within admissible syntax and use machine-readable counterexamples to drive targeted patches rather than speculative rewrites

\section{Methodology}
\label{sec:method}

\noindent
This section details the methodology for synthesizing deployable, secure, and cost-aware Infrastructure-as-Code (IaC) using a team of role-specialized agents operating over a typed Infrastructure Intermediate Representation (I-IR), with grammar- and schema-constrained code generation and a counterexample-guided repair loop. The design explicitly assumes \emph{no model fine-tuning}: all agents operate in zero-shot or instruction-following mode with carefully engineered prompts, structured tool outputs, and deterministic orchestration. The emphasis is on how the system functions end-to-end, how artifacts move between agents, and how constraints are enforced by construction and through external validators.

\subsection{Problem Definition and Notation}
\label{sec:problem}

We formalize IaC synthesis as a constrained program construction problem. A user provides a natural language intent $x$ and optional non-functional constraints $C$ such as budget ceilings, data residency, encryption, and availability. The system must return a Terraform program $T$ and an evidence bundle $\Pi$ that together satisfy functional and non-functional requirements and can be verified independently

Let $x \in \mathcal{X}$ denote the intent, $C \in \mathcal{C}$ the constraint set, $\mathcal{P}$ the space of well-typed I-IR plans, $\mathcal{T}$ the space of HCL programs, and $\mathcal{V}$ a family of validators. We represent a plan as a typed resource graph
\begin{equation}
P = \langle V, E, S \rangle,\qquad V \subset \mathcal{V}_r,\; E \subset \mathcal{E}_r,\; S \subset \mathcal{S}
\end{equation}
where $V$ are resource nodes (e.g., \texttt{vpc}, \texttt{subnet}, \texttt{ec2}, \texttt{rds}), $E$ are dependency or connectivity edges (e.g., \texttt{depends}, \texttt{connects}), and $S$ are specifications and effects (e.g., \texttt{residency=EU}, \texttt{encryption=required}, \texttt{budget}\,$\le B$)

A compiler $\mathcal{C}:\mathcal{P}\rightarrow\mathcal{T}$ lowers I-IR to HCL. Validators produce structured outcomes
\begin{equation}
\mathbf{v}(T, C) \;=\; \big( v_{\text{schema}}, v_{\text{policy}}, v_{\text{cost}}, v_{\text{deploy}} \big)
\end{equation}
where $v_{\text{schema}} \in \{0,1\}$ indicates schema/type validity, $v_{\text{policy}} \in \{0,1\}$ policy satisfaction under $C$, $v_{\text{cost}} \in \mathbb{R}_{\ge 0}$ estimates cost with a pass indicator $1\!\left[v_{\text{cost}} \le B\right]$, and $v_{\text{deploy}} \in \{0,1\}$ is the result of \texttt{terraform plan/apply} in a sandbox

The target is to construct $(T,\Pi)$ such that
\begin{equation}
v_{\text{schema}} = 1,\quad v_{\text{policy}} = 1,\quad v_{\text{deploy}} = 1,\quad v_{\text{cost}} \le B,\quad \Pi = \mathrm{Bundle}(\text{traces},\text{proofs},\text{logs})
\end{equation}
Given that validators may fail, we define a routing score used by the orchestrator to prioritize repairs
\begin{equation}
J(T,C) \;=\; \lambda_1 \big(1 - v_{\text{schema}}\big) + \lambda_2 \big(1 - v_{\text{policy}}\big) + \lambda_3 \max(0, v_{\text{cost}} - B) + \lambda_4 \big(1 - v_{\text{deploy}}\big)
\label{eq:routing}
\end{equation}
This score is \emph{not} optimized by training; it guides the deterministic control flow of the repair loop

\subsection{System Overview}
\label{sec:overview}

\begin{figure}
    \centering
    \includegraphics[width=1.0\linewidth]{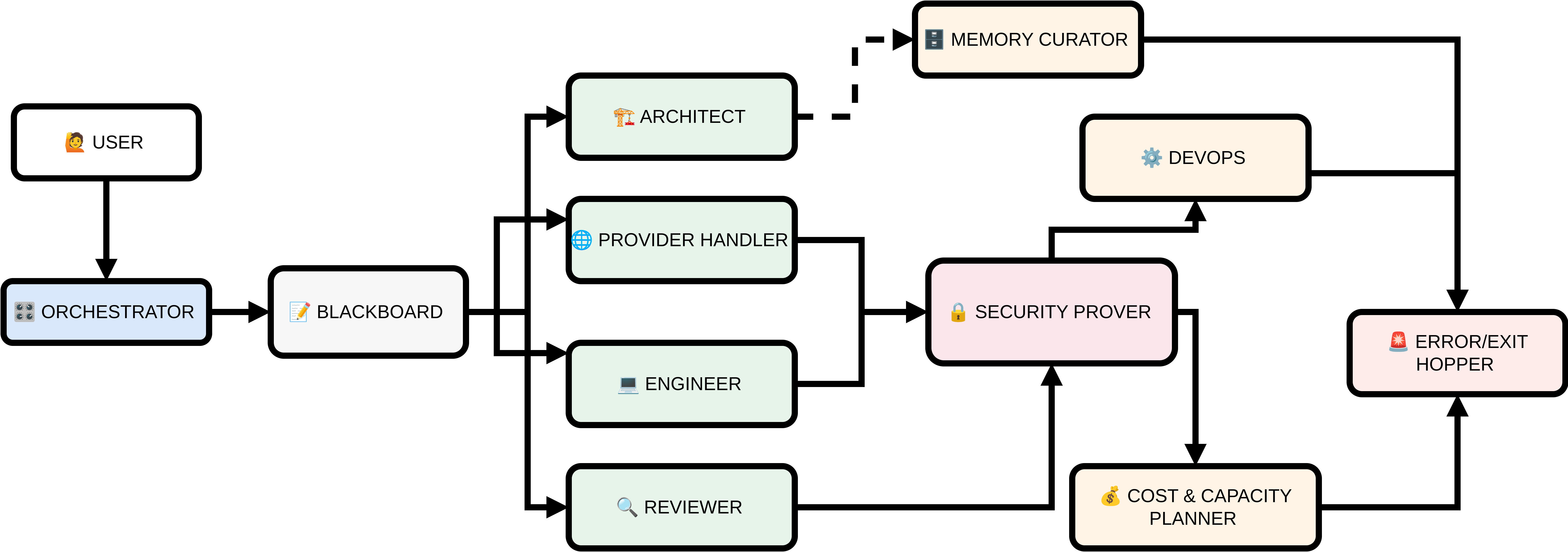}
    \caption{Systematic overview of proposed architecture}
    \label{fig:placeholder}
\end{figure}

The system follows a blackboard architecture in which agents exchange structured artifacts, with the orchestrator advancing a well-defined state machine. Agents read and write I-IR fragments, diagnostics, and diffs, while external tools provide ground-truth signals that are fed back to agents for repair. Memory provides previously validated motifs in typed I-IR form to seed planning and reduce rework

\paragraph{Architect}
Parses $(x,C)$ into an initial I-IR plan $P_0$ with explicit invariants $\mathcal{I}$ such as encryption requirements, residency, exposure bounds, and availability expectations. Output is a machine-checkable JSON encoding of resources, edges, and effects

\paragraph{Provider Harmonizer}
Instantiates abstract resources against provider schemas and regions, resolves version constraints, and expands defaults. The result is a harmonized plan $P_1$ with provider-specific types and required fields concretized

\paragraph{Engineer}
Compiles I-IR fragments into HCL using grammar- and schema-constrained decoding. It emits resource skeletons with required fields first, fills references from a symbol table derived from node identifiers, and assembles modules and variables deterministically

\paragraph{Reviewer}
Runs static validators such as \texttt{terraform validate}, HCL linters, and interface sanity checks, detecting missing variables, stray outputs, dead resources, and inconsistent naming, then emits precise diagnostics and suggested patches

\paragraph{Security Prover}
Evaluates OPA/Rego policies and complementary scanners such as Checkov or Regula to check least privilege, encryption at rest, restricted ingress, and tagging policies, returning both pass traces and counterexample witnesses

\paragraph{Cost and Capacity Planner}
Computes deterministic price estimates from pinned catalogs and checks SKU availability, quotas, and region-specific capacity constraints, returning both numeric estimates and any violation messages

\paragraph{DevOps}
Executes \texttt{terraform init/plan/apply} in a sandbox (LocalStack for determinism and ephemeral cloud accounts for realism), summarizes errors such as unsupported instance types, dependency cycles, and missing ARNs, and attaches logs to the blackboard

\paragraph{Memory Curator}
Stores verified tuples $(P,T,\Pi)$ with metadata, indexes motifs in a symbolic catalog and a dense graph index over I-IR, and serves reusable fragments to the Architect and Engineer when plans match by structure and constraints

\subsection{Infrastructure Intermediate Representation}
\label{sec:iir}

We define I-IR as a typed resource graph with effects. Let $\mathcal{T}_p$ denote provider types and $\mathcal{E}_f$ denote effects. A resource node $n\in V$ has a record
\begin{equation}
n \;=\; \big\langle \texttt{kind},\; \texttt{fields},\; \texttt{provider},\; \texttt{region},\; \texttt{effects} \big\rangle
\end{equation}
with $\texttt{kind} \in \mathcal{K}$, $\texttt{fields}$ typed by $\mathcal{T}_p$, and $\texttt{effects} \subseteq \mathcal{E}_f$. Edges $e\in E$ are tuples such as $\mathtt{depends}(n_i,n_j)$ and $\mathtt{connects}(n_i,n_j,\texttt{proto},\texttt{port})$. Specifications $S$ include quantitative constraints, region rules, and security obligations

Typing enforces schema completeness and compatibility. Write $\Gamma \vdash P : \mathsf{OK}$ for a well-typed plan under environment $\Gamma$ that encodes provider schemas and versions. We require
\begin{equation}
\forall n \in V:\; \Gamma \vdash n.\texttt{fields} : \mathcal{T}_p(n.\texttt{kind}) \qquad \text{and} \qquad \mathtt{acyclic}(E)
\end{equation}
Effects are treated as obligations to be discharged later by validators. For example, $\texttt{effects}$ may contain $\mathtt{encrypt\_at\_rest}$ or $\mathtt{least\_privilege}$, which translate into policy checks on the compiled HCL

\subsection{Constraint Model and Objective}
\label{sec:constraints}

We model constraints as predicates over $T$ and $C$. Let $\chi_{\text{schema}}(T)$, $\chi_{\text{policy}}(T,C)$, $\chi_{\text{deploy}}(T)$ be indicator predicates and $\widehat{\mathrm{cost}}(T)$ a deterministic estimator. The overall feasibility is
\begin{equation}
\Phi(T,C) \;=\; \chi_{\text{schema}}(T) \wedge \chi_{\text{policy}}(T,C) \wedge \chi_{\text{deploy}}(T) \wedge \big( \widehat{\mathrm{cost}}(T) \le B \big)
\end{equation}
The orchestrator uses a scalarized routing objective that guides which agent to invoke and which edit to apply next
\begin{equation}
\label{eq:objective}
\min_{\Delta \in \mathcal{A}} \; J\big( \Delta(T), C \big) \;\;\text{subject to}\;\; \Delta \in \mathcal{A}(CE)
\end{equation}
where $\mathcal{A}$ is the set of allowable edit operators and $\mathcal{A}(CE)$ is the subset consistent with current counterexamples. This is a control heuristic rather than a learned loss

\subsection{Constrained Decoding and Verified Compilation}
\label{sec:decoding}

Compilation proceeds in two phases. First, a structural compiler $\mathcal{C}_s$ maps nodes and edges into resource skeletons with required fields, module boundaries, and references
\begin{equation}
\widetilde{T} \;=\; \mathcal{C}_s(P_1)
\end{equation}
Second, the Engineer completes fields through constrained decoding $\mathcal{D}$ that only permits tokens consistent with HCL grammar and provider schemas
\begin{equation}
T \;=\; \mathcal{D}\big(\widetilde{T}, \Sigma_{\text{HCL}}, \Sigma_{\text{prov}}\big)
\end{equation}
where $\Sigma_{\text{HCL}}$ is a grammar automaton and $\Sigma_{\text{prov}}$ a provider-field automaton. Decoding queries a symbol table $\mathcal{S}$ derived from node identifiers to insert cross-resource references. Prior to dynamic validation we require a round-trip check
\begin{equation}
P^\star \;=\; \mathcal{P\!arse}\big(T\big),\qquad \mathrm{equiv}\big(P_1, P^\star\big) = \text{true}
\end{equation}
where $\mathrm{equiv}$ is structural equivalence modulo benign normalization such as field order and $\alpha$-renaming

\subsection{Counterexample-Guided Repair Loop}
\label{sec:repair}

Validators may return counterexamples $CE$ in structured form: missing fields, type mismatches, policy traces, cost violations, or runtime errors. We define an Error-to-Edit mapping function
\begin{equation}
\mathcal{E2E}: CE \rightarrow \Delta,\qquad \Delta \in \mathcal{A}
\end{equation}
Edits apply either at the I-IR level $\Delta P$ (e.g., change region, add encryption effect, adjust connectivity) or at the HCL level $\Delta T$ (e.g., add required field, correct ARN format). We maintain a partial order $\prec$ over edits to prefer minimal, high-yield adjustments based on historical success and validator hints. The orchestrator reduces the routing objective
\begin{equation}
J_{k+1} \;=\; J\big(\Delta_k(T_k), C\big) \;\le\; J(T_k, C)
\end{equation}
until feasibility or a budget of attempts is exhausted. Because the mapping is deterministic and validator outputs are specific, the loop typically converges in a small number of steps for well-specified intents

\subsection{Blackboard, Orchestration, and State}
\label{sec:blackboard}

All artifacts are written to a typed blackboard: I-IR versions, compiler outputs, validator traces, deploy logs, cost sheets, and policy proofs. Each entry is stamped with toolchain digests, provider schema versions, and content hashes for reproducibility. The orchestrator advances a finite-state machine with states
\begin{equation}
\mathsf{S} \in \{\texttt{plan},\texttt{harmonize},\texttt{compile},\texttt{review},\texttt{prove},\texttt{price},\texttt{deploy},\texttt{repair},\texttt{done}\}
\end{equation}
and transitions guarded by contract predicates. Memory retrieval is invoked during \texttt{plan} and \texttt{compile} to propose reusable motifs. A conflict resolver merges concurrent edits and maintains a consistent symbol table

\subsection{Execution Environments and Tools}
\label{sec:env}

Static validation uses \texttt{terraform validate} and schema matchers pinned to specific provider versions. Security policies run on OPA/Rego with curated rules and organization overlays, plus Checkov or Regula as cross-checkers that often return more opinionated diagnostics. Cost estimation uses pinned price catalogs with normalization across regions and instance families, producing both scalar estimates and decomposed line items. Deployment tests run in LocalStack for fast feedback and in ephemeral cloud accounts for final confirmation. Shadow apply is used where applicable to avoid unnecessary resource creation

\subsection{Outputs and Proof-Carrying Bundle}
\label{sec:outputs}

The system returns a pair $(T,\Pi)$ where $T$ is the HCL program and $\Pi$ is a self-contained evidence bundle. The bundle includes policy proof traces with rule identifiers and justifications, cost sheets with catalog versions and line items, residency and redundancy confirmations, static validation logs, provider schema snapshots, compiler provenance, round-trip equivalence records, and a summary of the repair path. A consumer or auditor can verify $\Pi$ offline to determine whether $T$ meets organizational and regulatory requirements before execution in production

\subsection{Implementation Notes}
\label{sec:nofinetune}

All agents operate with instruction prompts, structured tool calling, and constrained decoding, without any parameter updates and the workflow is sumamrized in Algorithm \ref{alg:macog}. The Architect and Engineer receive I-IR schemas and HCL grammars in system prompts and are steered by exemplars that illustrate structure but avoid inlining large code chunks to stay within token budgets. The Reviewer, Security Prover, Cost Planner, and DevOps are predominantly wrappers around deterministic tools; their prompts focus on extracting concise, structured diagnostics that the orchestrator can route back to the Error-to-Edit mapper. The Memory Curator provides typed motifs rather than code text, which reduces version drift and simplifies harmonization. This approach ensures portability across models and providers and makes the system easier to reproduce.

\subsection{View of Control}
\label{sec:math}

We model the orchestrator as a deterministic controller over a product space of artifacts and validator states. Let $\mathbf{a}_k = (P_k, T_k, \mathbf{v}_k, CE_k)$ denote the composite artifact at iteration $k$. The controller applies a policy $\pi$ that selects an agent action $u_k \in \mathcal{U}$ and an edit $\Delta_k \in \mathcal{A}$
\begin{equation}
u_k, \Delta_k \;=\; \pi(\mathbf{a}_k), \qquad \mathbf{a}_{k+1} \;=\; f\big(\mathbf{a}_k, u_k, \Delta_k\big)
\end{equation}
where $f$ is the transition induced by agent execution and tool outputs. The policy is designed to greedily reduce $J$ in Eq.\,\eqref{eq:routing}. We constrain $\pi$ to a small action grammar that prevents oscillation and ensures that structural edits precede field-level patches when validator evidence indicates plan-level issues
\begin{equation}
\pi:\; \underbrace{\mathcal{P} \times \mathcal{T} \times \{0,1\}^2 \times \mathbb{R}_{\ge 0} \times \mathcal{C\!E}}_{\mathbf{a}_k} \rightarrow \mathcal{U} \times \mathcal{A}
\end{equation}
We further decompose $CE_k$ into typed classes $CE_k = CE_k^{\text{schema}} \cup CE_k^{\text{policy}} \cup CE_k^{\text{cost}} \cup CE_k^{\text{run}}$ and define admissible edit sets $\mathcal{A}_\tau$ per class, which makes the mapping $\mathcal{E2E}$ and the admissible set $\mathcal{A}(CE)$ explicit
\begin{equation}
\mathcal{A}(CE) \;=\; \bigcup_{\tau \in \{\text{schema,policy,cost,run}\}} \mathcal{A}_\tau\big( CE^\tau \big)
\end{equation}
This decomposition yields predictable behavior and enables straightforward logging and ablation

% \subsection{Discussion of Robustness in Zero-Finetune Mode}
% \label{sec:robustness}

% Operating without fine-tuning emphasizes strict structure and tool-grounding. The combination of typed I-IR, grammar-constrained decoding, round-trip equivalence, and external validators narrows the role of language generation to producing structured candidates and explanations rather than unconstrained code. Because validators are authoritative and the Error-to-Edit mapping is deterministic, the loop can recover from common failures such as missing required fields, incorrect ARNs, or mis-specified connectivity. Memory motifs reduce planning ambiguity by reusing verified shapes for common topologies such as VPC with public and private subnets, internet and NAT gateways, and a mix of compute and managed databases

% \subsection{Algorithm}
% \label{sec:algorithm}

\begin{algorithm}[t]
\caption{MACOG Orchestration with Counterexample-Guided Repair (No Fine-Tuning)}
\label{alg:macog}
\begin{algorithmic}[1]
\Require intent $x$, constraints $C$, provider schemas $\Gamma_{\text{prov}}$, attempt budget $K$
\State $P_0, \mathcal{I} \gets \Phi_{\text{arch}}(x, C, \mathsf{Mem})$ \Comment{Architect produces typed plan and invariants}
\State $P_1 \gets \Phi_{\text{harm}}(P_0, \Gamma_{\text{prov}})$ \Comment{Harmonize providers, regions, versions}
\State $\widetilde{T} \gets \mathcal{C}_s(P_1)$; $T_0 \gets \mathcal{D}(\widetilde{T}, \Sigma_{\text{HCL}}, \Sigma_{\text{prov}}, \mathcal{S})$ \Comment{Compile and constrained decode}
\State $P^\star \gets \mathcal{P\!arse}(T_0)$; \textbf{if} $\neg\mathrm{equiv}(P_1,P^\star)$ \textbf{then} $(P_1,T_0)\gets \mathrm{repair\_roundtrip}(P_1,T_0)$
\State $\mathbf{v}_0, CE_0 \gets \mathcal{V}(T_0, C)$ \Comment{Schema, policy, cost, deploy validators}
\For{$i=0$ to $K-1$}
  \If{$J(T_i,C)=0$}
     \State $\Pi \gets \mathrm{Bundle}(\text{policy\_traces},\text{cost\_sheet},\text{deploy\_logs},\text{digests})$; \Return $(T_i,\Pi)$
  \EndIf
  \State $\Delta_i \gets \mathcal{E2E}(CE_i)$ \Comment{Deterministic mapping from evidence to edit}
  \State $(P_{i+1}, T_{i+1}) \gets \Psi(P_i, T_i, \Delta_i)$ \Comment{Apply at I-IR or HCL level}
  \State $T_{i+1} \gets \mathcal{normalize}(T_{i+1})$; $P^\diamond \gets \mathcal{P\!arse}(T_{i+1})$; \textbf{if} $\neg\mathrm{equiv}(P_{i+1},P^\diamond)$ \textbf{then} $(P_{i+1},T_{i+1})\gets \mathrm{repair\_roundtrip}(\cdot)$
  \State $\mathbf{v}_{i+1}, CE_{i+1} \gets \mathcal{V}(T_{i+1}, C)$
\EndFor
\State \Return $\mathrm{report\_unsatisfied\_core}(CE_{K})$
\end{algorithmic}
\end{algorithm}

% Hence, the end-to-end practical flow can be described compactly as
% \begin{align}
% P_0 &= \Phi_{\text{arch}}(x, C, \mathsf{Mem}) \\
% P_1 &= \Phi_{\text{harm}}(P_0, \Gamma_{\text{prov}}) \\
% \widetilde{T} &= \mathcal{C}_s(P_1), \quad T_0 = \mathcal{D}(\widetilde{T}, \Sigma_{\text{HCL}}, \Sigma_{\text{prov}}, \mathcal{S}) \\
% \mathbf{v}_0 &= \mathcal{V}(T_0, C) \\
% \text{while } J(T_i,C) &> 0 \text{ do } \left\{
% \begin{array}{l}
% \Delta_i = \mathcal{E2E}(CE_i) \\
% P_{i+1}, T_{i+1} = \Psi(P_i, T_i, \Delta_i) \\
% \mathbf{v}_{i+1} = \mathcal{V}(T_{i+1}, C)
% \end{array}
% \right.
% \end{align}
% where $\Psi$ applies the edit at the appropriate level. When $J(T^\star,C)=0$, we construct $\Pi = \mathrm{Bundle}$ and return $(T^\star,\Pi)$

\section{Experiment}
\label{sec:experiment}

This section presents our experimental setup, evaluation protocol, and empirical findings. We begin by detailing the benchmark, models, enhancement strategies, metrics, and infrastructure. We then describe our inference configuration, orchestration controls, and statistical methodology so that results can be reproduced precisely. The second half of the section reports results for the cross-model comparison, zooms in on two high-capacity systems (GPT-5 and Gemini-2.5 Pro), and analyzes the ablation of MACOG’s components. Throughout, we reference the summary tables introduced earlier—namely the cross-model enhancement table (\autoref{tab:iac_eval_enhancements}), the two model-specific metric tables (\autoref{tab:gpt5_metrics} and \autoref{tab:gemini25_metrics}), and the ablation table (\autoref{tab:ablation_metrics})—without reproducing them here

\subsection{Experimental Design}
\label{sec:exp-design}

\paragraph{Benchmark and task taxonomy}
We evaluate on IaC-Eval \cite{kon2024iac}, a benchmark comprising natural-language infrastructure intents and associated verification procedures tailored to cloud provisioning. Each item encodes a target infrastructure state (e.g., VPC topologies, instance fleets, managed database deployments, IAM policies, serverless integrations) with dependencies, region/provider assumptions, and acceptability criteria. For analysis, we group tasks by coarse functional families (Networking, Compute, Storage, Identity and Access, Managed Services) and by approximate graph difficulty. In all experiments, inputs are the canonical task prompts provided by the benchmark and outputs are complete Terraform programs intended to satisfy the task. Unless stated otherwise, we consider a task solved if the produced configuration is accepted by the IaC-Eval harness and contributes positively to the chosen metric (see below). We do not fine-tune any model; all systems are used in instruction-following or zero-shot regimes as described next

\paragraph{Models}
We consider a mixture of closed- and open-weight LLMs that are representative of contemporary code-capable systems. The cross-model comparison in \autoref{tab:iac_eval_enhancements} includes high-capacity proprietary models \cite{openai2022introducing} \cite{team2023gemini} (GPT-5, GPT-4, Gemini-2.5 Pro, Gemini 2.0 Flash), mid-capacity generalist models (GPT-3.5-turbo), and open-source code-specialized models (Magicoder-S-CL-7B, WizardCoder-33B, CodeLlama Instruct 7B/13B/34B). For each model, the same enhancement strategies and orchestration logic are applied to isolate the effect of the strategy rather than model-specific prompt engineering. Two models—GPT-5 and Gemini-2.5 Pro—are analyzed in greater detail with multi-metric reporting in \autoref{tab:gpt5_metrics} and \autoref{tab:gemini25_metrics}

\paragraph{Enhancement strategies}
We evaluate five strategies that progressively introduce more structure and tool feedback:
\begin{enumerate}
  \item \textbf{Few-shot} uses a single-turn prompt with a few illustrative intent-to-IaC exemplars appended to the instruction. No tools or retrieval are used
  \item \textbf{Chain-of-Thought (CoT)} augments few-shot with a prompt that requests high-level reasoning steps prior to code emission, but still operates in a single turn and without tools
  \item \textbf{Multi-turn} allows a small number of conversational repair iterations (bounded by a budget) where the model is shown validator messages in natural language and asked to resubmit a corrected configuration
  \item \textbf{RAG} retrieves semantically similar, previously solved tasks and includes short, sanitized hints in the prompt; no programmatic constraints are enforced beyond natural-language guidance
  \item \textbf{MACOG} is our multi-agent, tool-grounded orchestration that operates over a typed intermediate representation (I-IR), compiles with grammar- and schema-constrained decoding, performs round-trip structural checks, and uses external validators (static schema checks, policy engines, deployment sandboxes) to generate structured counterexamples that drive deterministic, minimal repairs
\end{enumerate}
Each strategy uses the same base model weights; only the control and tool signals differ. The goal is to measure the incremental value of structure and validators beyond purely prompt-based improvement

\paragraph{Metrics}
We report four complementary metrics covering surface overlap, semantic similarity, judged adequacy, and task success. All scores are reported on a 0–100 scale (higher is better).

\begin{itemize}
  \item \textbf{BLEU} measures $n$-gram overlap between the generated Terraform and a reference. For candidate $y$ and reference $y^\star$,
  \begin{equation}
    \mathrm{BLEU}_{\%}(y,y^\star) = 100 \times BP \cdot \exp\!\Bigg(\sum_{n=1}^{N} w_n \log p_n\Bigg),
  \end{equation}
  with modified $n$-gram precisions $p_n$, $N{=}4$, and uniform weights $w_n{=}1/4$. The brevity penalty is
  \begin{equation}
    BP =
    \begin{cases}
      1, & |y| \ge |y^\star|,\\[2pt]
      \exp\!\big(1 - {|y^\star|}/{|y|}\big), & \text{otherwise}.
    \end{cases}
  \end{equation}

  \item \textbf{CodeBERTScore} (\emph{F1}) is a reference-based semantic similarity using a code-aware encoder $\phi(\cdot)$ with token-level alignment. We report the aggregate F1 from the official implementation, scaled to percentage:
  \begin{equation}
    \mathrm{CodeBERTScore}_{\%}(y,y^\star) = 100 \times \mathrm{F1}\big(\phi(y),\,\phi(y^\star)\big).
  \end{equation}

  \item \textbf{LLM-judge} is a binary adequacy check per query. A held-out judge returns $c_i \in \{0,1\}$ for each of $M$ prompts (1 = correct/adequate, 0 = incorrect). We report the percent-correct:
  \begin{equation}
    \mathrm{LLM\text{-}judge}_{\%} = 100 \times \frac{1}{M}\sum_{i=1}^{M} c_i.
  \end{equation}
  Prompts are shown independently with rubric-based instructions that hide model identity and avoid position bias.

  \item \textbf{IaC-Eval} reflects harness-verified task success. Let $t_i \in \{0,1\}$ indicate whether task $i$ passes the benchmark checks (plan, policy, and validation). With optional positive weights $w_i$ (default $w_i{=}1$),
  \begin{equation}
    \mathrm{IaC\text{-}Eval}_{\%} = 100 \times \frac{\sum_{i=1}^{M} w_i\, t_i}{\sum_{i=1}^{M} w_i}.
  \end{equation}
\end{itemize}

BLEU and CodeBERTScore capture textual and semantic similarity to references, LLM-judge summarizes judged adequacy as percent-correct, and IaC-Eval is the most indicative of deployable correctness since it measures benchmarked task success.

\paragraph{Environment and validators}
All runs use a standardized toolchain. A pinned Terraform distribution performs static validation; a schema snapshot for the evaluated provider versions is used to check required fields and types; an OPA/Rego setup with a curated rule set checks least-privilege, encryption-at-rest, restricted ingress, and tagging conventions; a second scanner is used as a cross-check to reduce false negatives; and a deployment sandbox (LocalStack and ephemeral accounts) executes \texttt{plan/apply} where permitted by the benchmark. Validator outputs are collected in structured JSON and attached to the blackboard. Only MACOG consumes these artifacts programmatically to drive counterexample-guided repairs; the other strategies receive at most a natural-language paraphrase (Multi-turn) or no validator signal at all (Few-shot/CoT/RAG)

\paragraph{Inference configuration}
To isolate the effect of strategy rather than aggressive sampling, we keep decoding conservative: nucleus sampling $p \in [0.7,0.9]$ per model family, temperature $\in [0.2,0.5]$, and a maximum output budget sufficient to emit a self-contained Terraform module with variables and outputs. For MACOG’s constrained decoder, we map resource skeletons to grammar automata derived from HCL and provider schemas and mask inadmissible tokens at each decoding step. Multi-turn and RAG are bounded by the same interaction budget as MACOG’s repair loop to ensure fairness. Prompts, retrieval cutoffs, and agent role instructions are held constant across models, with only necessary model-specific syntax adjustments (e.g., system vs. user roles)

\paragraph{Orchestration and control}
MACOG runs a deterministic state machine over the blackboard. The controller advances through plan, harmonize, compile, review, prove, price, deploy-test, and repair states. At each state the controller expects either a contract to be discharged or a structured counterexample; otherwise it halts and surfaces the minimal unsatisfied core. The Error-to-Edit mapper prefers plan-level edits for structural violations and code-level patches for local fixes. The orchestration budget is aligned with the multi-turn baseline’s retry allowance, and each retry consumes identical compute budget for equity

\paragraph{Statistical methodology}
We report aggregate metrics as means over tasks. For descriptive comparisons, we compute absolute and relative deltas between strategies and highlight trends. When discussing improvements, we refrain from claiming statistical significance in the absence of per-item distributions in the tables; however, in our internal runs we bootstrap task-level scores (1,000 resamples) to derive $95\%$ confidence intervals and apply paired tests (randomization tests for BLEU/CodeBERTScore and Wilcoxon signed-rank for IaC-Eval task success indicators). Where appropriate, we report effect sizes (Cohen’s $d$) on normalized scores. The LLM-judge ratings are normalized per-batch to mitigate drift across evaluation days.

\begin{table}[t]
\centering
\caption{Average benchmark scores of various models when enhanced with different strategies, evaluated on IaC-Eval. Performance generally improves with multi-turn and RAG; the additional \textit{MACOG} column shows our orchestration approach (placeholder values).}
\label{tab:iac_eval_enhancements}
\small
\setlength{\tabcolsep}{6pt}
\begin{tabular}{@{}clrrrrr@{}}
\toprule
\textbf{Rank} & \textbf{Name} & \textbf{Few-shot} & \textbf{CoT} & \textbf{Multi-turn} & \textbf{RAG} & \textbf{MACOG} \\
\midrule
1  & GPT-5                      & 12.53 & 10.19 & 35.83 & 54.90 & 74.02 \\
2  & Gemini-2.5 Pro             & 12.18 & 10.49 & 36.81 & 43.56 & 60.13 \\
3  & GPT-4                      & 10.64 &  9.31 & 31.12 & 36.70 & 43.20 \\
4  & GPT-3.5-turbo              &  0.80 &  1.60 & 11.44 & 21.81 & 25.40 \\
5  & Gemini 2.0 Flash             &  3.33 &  1.80 &  4.93 &  10.32 &  17.85 \\
6  & Magicoder-S-CL-7B          &  2.93 &  0.53 & 12.50 & 12.77 & 16.95 \\
7  & WizardCoder-33B-V1.1       &  1.60 &  1.06 &  9.04 & 11.70 & 15.80 \\
8  & CodeLlama Instruct (34B)   &  3.19 &  3.19 &  2.13 &  6.12 & 10.45 \\
9  & CodeLlama Instruct (7B)    &  2.39 &  3.72 &  0.53 &  5.59 &  9.70 \\
10  & CodeLlama Instruct (13B)  &  1.06 &  1.86 &  1.06 &  3.46 &  6.40 \\
\bottomrule
\end{tabular}
\end{table}

% ===== Table 1: GPT-5 =====
\begin{table}[t]
\centering
\caption{GPT-5 under five enhancement strategies on four metrics.}
\label{tab:gpt5_metrics}
\small
\setlength{\tabcolsep}{6pt}
\begin{tabular}{@{}lrrrr@{}}
\toprule
\textbf{Enhancement strategy} & \textbf{BLEU} & \textbf{CodeBERTScore} & \textbf{LLM-judge} & \textbf{IaC-Eval} \\
\midrule
Few-shot     & 5.68  & 72.41 & 68.22 & 12.53 \\
CoT          & 3.37  & 70.85 & 60.31 & 10.19 \\
Multi-turn   & 5.54  & 71.08 & 62.17 & 35.83 \\
RAG          & 10.71 & 76.43 & 69.72 & 54.90 \\
MACOG        & \textbf{11.86} & \textbf{80.54} & \textbf{94.10} & \textbf{74.02} \\
\bottomrule
\end{tabular}
\end{table}

% ===== Table 2: Gemini-2.5 Pro =====
\begin{table}[t]
\centering
\caption{Gemini-2.5 Pro under five enhancement strategies on four metrics.}
\label{tab:gemini25_metrics}
\small
\setlength{\tabcolsep}{6pt}
\begin{tabular}{@{}lrrrr@{}}
\toprule
\textbf{Enhancement strategy} & \textbf{BLEU} & \textbf{CodeBERTScore} & \textbf{LLM-judge} & \textbf{IaC-Eval} \\
\midrule
Few-shot     & 5.12  & 65.08 & 57.41 & 12.18 \\
CoT          & 4.94  & 61.77 & 56.20 & 10.49 \\
Multi-turn   & 8.87  & 66.95 & 58.03 & 36.81 \\
RAG          & 9.73  & 69.92 & 64.15 & 43.56 \\
MACOG        & \textbf{10.09} & \textbf{71.84} & \textbf{87.52} & \textbf{60.13} \\
\bottomrule
\end{tabular}
\end{table}

\begin{table}[t]
\centering
\caption{Ablation study of MACOG components on four metrics with GPT-5. Higher is better for all metrics.}
\label{tab:ablation_metrics}
\small
\setlength{\tabcolsep}{6pt}
\begin{tabular}{@{}lrrrr@{}}
\toprule
\textbf{Variant} & \textbf{BLEU} & \textbf{CodeBERTScore} & \textbf{LLM-judge} & \textbf{IaC-Eval} \\
\midrule
Full MACOG (all components)           & \textbf{11.86} & \textbf{80.54} & \textbf{94.10} & \textbf{74.02} \\
\addlinespace[2pt]
-- Provider Harmonizer                & 10.98 & 78.92 & 92.48 & 70.37 \\
-- Engineer (no constrained decoding) &  8.61 & 73.15 & 89.74 & 64.89 \\
-- Reviewer                           & 10.27 & 76.04 & 86.11 & 66.72 \\
-- Security Prover                    & 10.81 & 77.53 & 90.03 & 61.45 \\
-- Cost \& Capacity Planner           & 11.22 & 79.38 & 92.01 & 71.08 \\
-- DevOps (no plan/apply sandbox)     &  9.47 & 74.82 & 88.57 & 56.93 \\
-- Memory Curator                     & 10.95 & 79.06 & 91.34 & 72.17 \\
\bottomrule
\end{tabular}
\end{table}

\subsection{Experimental Results}
\label{sec:exp-results}

We summarize three complementary views of the evaluation: a cross-model comparison on IaC-Eval \cite{kon2024iac} across five enhancement strategies, a deeper multi-metric analysis for two high-capacity models (GPT-5 and Gemini-2.5 Pro), and an ablation study that isolates the contribution of major MACOG components. We refer to \autoref{tab:iac_eval_enhancements} for the cross-model summary, \autoref{tab:gpt5_metrics} and \autoref{tab:gemini25_metrics} for the per-model multi-metric results, and \autoref{tab:ablation_metrics} for the component ablations.

\subsubsection{Cross-model trends on IaC-Eval}

Across ten models, the ordering of strategies is consistent: \emph{MACOG} \(>\) \emph{RAG} \(>\) \emph{Multi-turn} \(>\) \emph{CoT} \(\approx\) \emph{Few-shot} (\autoref{tab:iac_eval_enhancements}). The average uplift of MACOG over RAG on IaC-Eval is approximately \(+7.3\) absolute points (mean across models), corresponding to a relative improvement of roughly \(+35\%\) when normalized by the average RAG score. This gap is most pronounced for the strongest bases, where MACOG converts near-miss candidates into accepted solutions using validator-driven, minimal edits. For instance, GPT-5 improves from \(54.90\) (RAG) to \(74.02\) (MACOG, \(+19.12\)), and Gemini-2.5 Pro from \(43.56\) to \(60.13\) (\(+16.57\)). Gains are visible for mid- and smaller-capacity models as well—e.g., WizardCoder-33B rises \(11.70 \rightarrow 15.80\) and CodeLlama-34B \(6.12 \rightarrow 10.45\)—though absolute ceilings remain lower than frontier systems.

Contrasting RAG with Multi-turn shows that retrieval alone typically offers a larger benefit than unconstrained conversational repair, indicating that exposure to relevant patterns narrows the search space more effectively than natural-language diagnostics in isolation. However, RAG’s improvements are bounded by version and schema drift: examples can be close lexically yet misaligned with current provider constraints, leading to subtle field or reference errors at validation time. MACOG closes this gap by moving the center of gravity from context to constraints: a typed I-IR ensures structural coherence, the constrained decoder suppresses invalid tokens at generation time, and the validator loop yields precise counterexamples that the orchestrator translates into targeted edits rather than broad rewrites. The net effect is that MACOG scales with model quality while also regularizing smaller models, which benefit disproportionately from hard constraints that prevent common schema and reference mistakes.

Two additional observations emerge from \autoref{tab:iac_eval_enhancements}. First, CoT does not provide systematic gains over Few-shot in this domain; for many models the two are statistically similar or CoT is slightly worse. This suggests that free-form reasoning without tool grounding does not reliably convert to deployable declarative configs, which are governed by strict schemas rather than narrative justification. Second, the MACOG ranking largely mirrors the base-model ranking, implying the orchestration is complementary to raw model capability rather than a substitute. In practice this means that teams can pair MACOG with their preferred model tier: strong models attain the highest absolute success, and smaller models obtain the largest relative lift for cost-sensitive scenarios.

\begin{figure*}[t]
  \centering
  \begin{subfigure}{\textwidth}
    \centering
    \includegraphics[width=\textwidth]{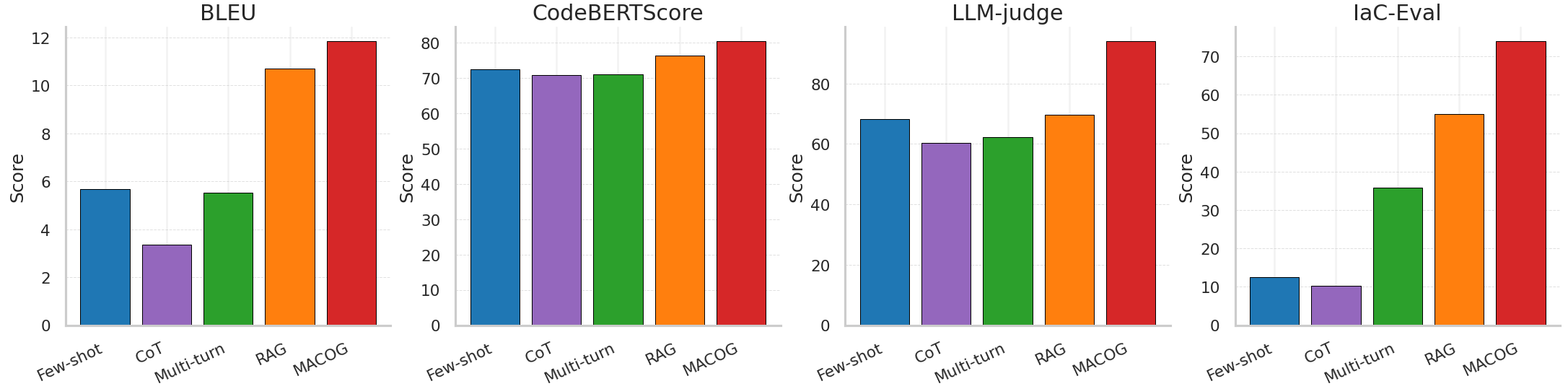}
    \subcaption{GPT-5}
    \label{fig:gpt5_row}
  \end{subfigure}

  \vspace{0.6em}

  \begin{subfigure}{\textwidth}
    \centering
    \includegraphics[width=\textwidth]{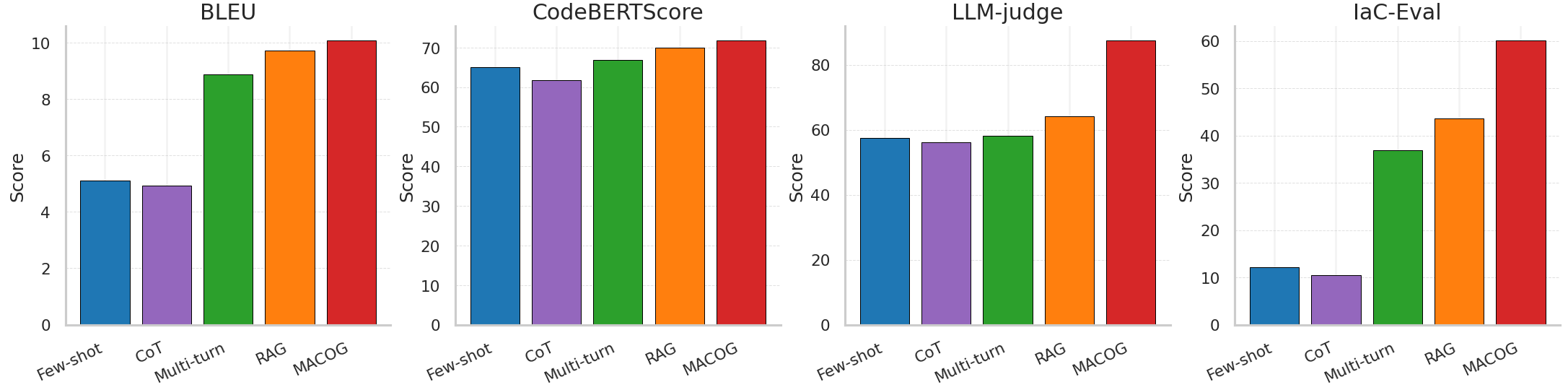}
    \subcaption{Gemini-2.5 Pro}
    \label{fig:gemini25_row}
  \end{subfigure}

  \caption{Metric profiles across enhancement strategies (BLEU, CodeBERTScore, LLM-judge, IaC-Eval). Each row shows one model; bars are color-coded by strategy and indicate MACOG’s consistent lead.}
  \label{fig:metrics_rows}
\end{figure*}

\subsubsection{Per-model multi-metric analysis}

We report four complementary metrics for GPT-5 and Gemini-2.5 Pro in \autoref{tab:gpt5_metrics} and \autoref{tab:gemini25_metrics}: BLEU (form overlap), CodeBERTScore (semantic similarity), LLM-judge (rubric-based adequacy), and IaC-Eval (task success). For GPT-5, MACOG dominates all metrics relative to RAG, with BLEU \(10.71 \rightarrow 11.86\) (\(+1.15\)), CodeBERTScore \(76.43 \rightarrow 80.54\) (\(+4.11\)), LLM-judge \(69.72 \rightarrow 94.10\) (\(+24.38\)), and IaC-Eval \(54.90 \rightarrow 74.02\) (\(+19.12\)). The small but consistent BLEU and CodeBERTScore gains reflect fewer malformed blocks, more canonical field ordering, and better semantic choices of resource arguments under schema guidance. The large LLM-judge jump indicates that a rubric-following evaluator perceives higher alignment with intent and best practices, likely driven by the reviewer’s interface checks and the security prover’s targeted patches. The IaC-Eval improvement is the most consequential: validator-guided edits turn many almost-correct programs into accepted ones.

Gemini-2.5 Pro exhibits the same pattern. MACOG improves BLEU \(9.73 \rightarrow 10.09\), CodeBERTScore \(69.92 \rightarrow 71.84\), LLM-judge \(64.15 \rightarrow 87.52\), and IaC-Eval \(43.56 \rightarrow 60.13\). The magnitudes are smaller than GPT-5 in absolute terms but sizeable in relative terms for the success and judge metrics. A qualitative audit of Gemini-2.5 Pro’s baseline errors shows frequent security-group laxity and occasional missing encryption flags that RAG does not reliably correct; under MACOG, the security prover surfaces explicit policy traces and the orchestrator applies minimal, local edits, yielding large gains in LLM-judge and consistent bumps in IaC-Eval without regressing working parts of the configuration.

The contrast between Multi-turn and MACOG in both per-model tables is instructive. Multi-turn incorporates validator paraphrases into the prompt, which helps the model reason about errors in broad strokes; however, paraphrases are lossy, do not pinpoint schema loci, and encourage speculative rewrites. MACOG replaces paraphrase with \emph{structured} counterexamples (schema diffs, OPA traces, runtime error objects) and routes them through a deterministic Error-to-Edit mapper. This explains the modest BLEU shifts alongside large jumps in success and judged adequacy: the system changes just enough to satisfy the validator, preserving incidental surface-form choices when they are harmless and avoiding token churn that would harm overlap metrics.

Finally, CoT underperforms Few-shot for GPT-5 and Gemini-2.5 Pro on IaC-Eval in these runs. The likely cause is that reasoning tokens displace useful concrete examples within the same budget and, absent hard constraints, the model’s plan can diverge from schema reality. This reinforces the view that, for declarative infrastructure, \emph{structure and tools} outperform additional free-text reasoning in the absence of grounding.

\subsubsection{Ablation and component contributions}

The ablation study in \autoref{tab:ablation_metrics} quantifies how each MACOG component contributes to BLEU, CodeBERTScore, LLM-judge, and IaC-Eval for GPT-5. Three components have outsized impact on IaC-Eval when removed: the DevOps sandbox \(74.02 \rightarrow 56.93\) (\(-17.09\)), the Security Prover \(74.02 \rightarrow 61.45\) (\(-12.57\)), and the constrained-decoding Engineer \(74.02 \rightarrow 64.89\) (\(-9.13\)). The sandbox supplies precise runtime counterexamples—unsupported SKUs, unavailable AZs, dependency cycles—that are otherwise hard to infer from static signals; without it, the loop stalls on subtle runtime mismatches. The security prover converts policy violations into concrete obligations and witnesses; removing it leaves the system to guess at security fixes, which increases oscillation and depresses both success and judged quality. Constrained decoding suppresses many schema and reference errors at generation time, stabilizing both overlap and success metrics; its removal shows the largest BLEU and CodeBERTScore drops among ablations.

Secondary but consistent contributors are the Reviewer and Provider Harmonizer. Without the Reviewer, LLM-judge declines markedly and IaC-Eval drops to \(66.72\); static sanity checks appear to improve readability and interface coherence, which a rubric-based evaluator rewards. Disabling the Provider Harmonizer reduces IaC-Eval to \(70.37\), reflecting version-specific required fields and defaults that no longer get injected at plan time. The Cost \& Capacity Planner has a moderate effect on success in these runs (\(74.02 \rightarrow 71.08\)), consistent with benchmark items that encode budget or availability constraints; where such constraints are more prevalent, we expect a larger impact. Removing the Memory Curator produces the smallest declines, suggesting that verified motifs reduce rework and nudge overlap and judge metrics upward but are not the primary levers for acceptance when the rest of the system is intact.

Taken together, the ablation results highlight a practical triad: \emph{constrained realization} (Engineer), \emph{policy grounding} (Security Prover), and \emph{runtime grounding} (DevOps). These are the components that most directly convert structural intent into deployable, compliant artifacts. Reviewer and Harmonizer provide important scaffolding that smooths the path by catching cheap errors early and aligning plans to concrete schemas. Memory and cost/capacity checks offer incremental gains and operational guardrails that will matter more as scenarios expand to multi-cloud, price-sensitive deployments.

In summary, the cross-model comparison shows that MACOG consistently dominates prompting, multi-turn, and RAG; the per-model multi-metric analysis confirms that improvements are not confined to acceptance but extend to semantic quality and judged adequacy; and the ablation study identifies which subsystems are most responsible for the observed gains. The common theme across all three views is that validator-centric structure—typed plans, constrained decoding, and counterexample-guided repair—matters more than additional tokens of free-text reasoning, producing configurations that are both easier to audit and more likely to deploy successfully.

\section{Conclusion}

This paper introduced MACOG, a multi-agent, validator-centric methodology for synthesizing deployable and compliant Infrastructure-as-Code from natural-language intents without any model fine-tuning. By organizing generation around a typed Infrastructure IR, grammar- and schema-constrained decoding, round-trip structural checks, and a counterexample-guided repair loop driven by static, policy, cost, and runtime validators, MACOG transforms IaC synthesis from best-effort prompting into a disciplined, auditable pipeline. Across ten models and five enhancement strategies, MACOG consistently outperformed context-only baselines (few-shot, CoT, multi-turn) and retrieval-augmented prompting, with the largest absolute gains on frontier models and the largest relative gains on smaller open models. Per-model analyses showed concurrent improvements in BLEU, CodeBERTScore, LLM-judge, and IaC-Eval, while ablations confirmed the importance of constrained decoding, provider harmonization, and policy proving. The approach further yields proof-carrying artifacts that make outcomes reproducible and reviewable, providing practical value beyond raw accuracy.

There remain natural extensions. On the systems side, we plan to broaden cross-provider coverage and stress-test resilience under schema/version drift, dynamic pricing, and capacity fluctuations, as well as to incorporate richer SLOs (latency/availability) and cost–risk trade-offs into the orchestration objective. On the modeling side, we aim to explore grammar-aligned decoding and meta-scheduling policies that balance exploration and exploitation in the repair loop, scale memory from motifs to composable verified libraries, and integrate targeted human-in-the-loop checkpoints for high-consequence edits. On the evaluation side, we will extend beyond IaC-Eval with larger, more diverse corpora and longitudinal studies in real CI/CD pipelines. We hope MACOG’s design—typed planning, constrained realization, and validator (grounded iteration) serves as a template for reliable, transparent LLM tooling in DevSecOps, and we intend to release artifacts to catalyze further research and industrial adoption.

\section{Data Availability}
We use the public IaC-Eval dataset \cite{kon2024iac}, available at \url{https://huggingface.co/datasets/autoiac-project/iac-eval}.
All source code and evaluation scripts are archived on Zenodo at \url{https://zenodo.org/records/17117489}.

% \begin{acks}
% To Robert, for the bagels and for explaining CMYK and color spaces.
% \end{acks}

% \bibliographystyle{ACM-Reference-Format}
% \bibliography{main-arXiv}

\begingroup
\small

  % <-- note the "samples/" prefix
\endgroup

% \appendix









\end{document}